%% file: DIRECTFN_quad.tex
\documentclass[journal]{IEEEtran}

\usepackage{amsmath}
\usepackage{subfigure}
\usepackage{bm}
\usepackage[noadjust]{cite}

\usepackage{graphicx}
\graphicspath{{figures/}}

\renewcommand{\[}{\begin{equation}}
\renewcommand{\]}{\end{equation}}

\renewcommand{\{}{\begin{eqnarray}}
\renewcommand{\}}{\end{eqnarray}}

\renewcommand{\vec}{\mathbf}

\title{On the Generalization of DIRECTFN for Singular Integrals Over Quadrilateral Patches}
\author{Alexandra Tambova, Mikhail Litsarev, Georgy Guryev, and Athanasios G. Polimeridis~\IEEEmembership{Senior Member,~IEEE}
	\thanks{The authors are with the Center for Computational Data-Intensive Science and Engineering, Skoltech, Moscow, Russia (e-mail: a.polimeridis@skoltech.ru).}}

\begin{document}
\maketitle

\begin{abstract}
	A set of fully numerical algorithms for evaluating the four-dimensional singular integrals arising from Galerkin surface integral equation methods over conforming quadrilateral meshes is presented. This work is an extension of DIRECTFN, which was recently developed for the case of triangular patches, utilizing in a same fashion a series of coordinate transformations together with appropriate integration re-orderings. The resulting formulas consist of sufficiently smooth kernels and exhibit several favorable characteristics when compared with the vast majority of the methods currently available. More specifically, they can be applied---without modifications---to the following challenging cases: 1) weakly and strongly singular kernels, 2) basis and testing functions of arbitrary order, 3) planar and curvilinear patches, 4) problem-specific Green functions (e.g.  expressed in spectral integral form), 5) spectral convergence to machine precision. Finally, we show that the overall performance of the fully numerical schemes can be further improved by a judicious choice of the integration order for each dimension.
\end{abstract}


\input{introduction}

\input{formulation}

\input{results}

\input{conclusion}

\section*{Acknowledgments}
This work was supported in part by grants from the Skoltech-MIT Next Generation Program.

\bibliography{IEEEabrv,References}
\bibliographystyle{IEEEtran}
\end{document}

%% file: introduction.tex
\section{INTRODUCTION}

The evaluation of multi-dimensional singular integrals arising from electromagnetic surface integral equation formulations has been under scrutiny since the very early days of computer-aided analysis of scattering and radiation phenomena involving complex geometries~\cite{Harrington_book}. Indeed, there is a plethora of numerical techniques especially designed for the accurate and efficient computation of these integrals, that can be roughly categorized into two main groups, the singularity cancellation~\cite{Duffy1982,Telles1987,Graglia1987,Schwab1992,Klees1996,Rossi1999,Herschlein2002,Cai2002,Jorgensen2004,Khayat2005,Tang2006,Tong2007,Ismatullah2008,Fink2008,Graglia2008,Ding2009,Yuan2009,Asvestas2010,Zhu2011,Kaur2011,Vipiana2013,Botha2013} and the singularity subtraction~\cite{Wilton1984,Caorsi1993,Graglia1993,Eibert1995,Notaros1997,Bluck1997,Arcioni1997,Hodges1997,Oijala2003,Jarvenpaa2003,Jarvenpaa2006,Hanninen2006,Notaros2008} methods. Of course this is by no means an exhaustive list of the various contributions developed over a span of four decades, but they represent the two main schools of thought.

Despite their different philosophy, both singularity cancellation and singularity subtraction methods share an important common characteristic, the regularization of the singular potential integrals, i.e. the inner 2-D integrals of the original 4-D Galerkin inner products. As it was shown recently, though, there are certain advantages in considering the complete 4-D integrals, especially in the case of strongly singular kernels or when high accuracy is needed. More specifically, a new class of semi-analytical methods was developed where the singularity cancellation approach was generalized with the help of analytical integrations and appropriate interchanges in the order of the associated one-dimensional integrations \cite{Taylor2003,Polimeridis2008,Polimeridis2010c,Polimeridis2011,Polimeridis2011b,DEMCEM,Reid2015}, hence resulting in sufficiently smooth integrals of reduced dimensionality that can be easily computed via simple Gaussian integration. 

However,  the pertinent analytical integrations have a direct impact on the versatility of these semi-analytical methods, excluding from their repertoire some interesting cases, e.g., singular integrals over curvilinear elements, strongly singular integrals that arise from analytical shape derivatives over planar elements \cite{Kataja2013}, and kernels with Green functions expressed in spectral form. In the course of recent investigations, it was discovered that the semi-analytical integrations of the above mentioned methods render them more efficient but they do not contribute in the further regularization of the kernels. Indeed, the series of coordinate transformations together with the integration re-orderings would suffice to produce smooth kernels. These new findings led to the development of a fully numerical method, dubbed DIRECTFN, which preserves the convergence properties of the semi-analytical schemes while extending significantly their range of applicability \cite{Polimeridis2013}.

Arguably, the vast majority of the numerical methods mentioned above were developed for evaluating singular integrals over triangles, mainly due to the profound impact of the celebrated paper by Rao, Wilton, and Glisson~\cite{Rao1982} on the computational electromagnetics community, and the flexibility the triangular tesselations offer in modeling arbitrary geometries. However, modern computer-aided design software enables the analysis of complex geometries in terms of flat or curvilinear quadrilateral patches, which can describe just as accurately the geometry with far fewer degrees of freedom~\cite{Kolundzija_book,Kolundzija1982,Deshpande,Djordjevic2004,WIPLD}. In addition, similar cases might arise from the dimensionality reduction of 6-D integrals over polyhedral elements, as shown in recently developed volume integral equation methods~\cite{Polimeridis2013b,Polimeridis2014}. To the best of our knowledge, there are only a handful of papers in the literature dedicated to the evaluation of singular integrals over flat and curvilinear quadrilateral patches~\cite{Djordjevic2004,Jorgensen2004,Notaros2008,Ding2009,Yuan2009,Manich2014}, and it is quite clear that they haven't reached the performance levels of those for triangular patches.

The primary objective of this paper is the extension of the DIRECTFN method to the case of 4-D singular integrals over quadrilateral patches. As it is shown in the following, DIRECTFN, unlike standard singularity subtraction and cancellation methods, requires a series of complicated algebraic manipulations and its extension from triangular to quadrilateral domains is by no means trivial. This work concludes our research program on the evaluation of singular Galerkin inner products for surface integral equation methods and offers---in combination with the original DIRECTFN paper---a general framework that can seamlessly cover triangular and quadrilateral integration domains for the following challenging cases:
\begin{itemize}
\item weakly and strongly singular kernels
\item basis and testing functions of arbitrary order
\item planar and curvilinear patches
\item problem-specific Green functions (e.g.  expressed in spectral integral form)
\item spectral convergence to machine precision
\end{itemize}
It is also worth noting that the fully numerical schemes presented herein, unlike other semi-analytical methods~\cite{Taylor2003,Polimeridis2008,Polimeridis2010c,Polimeridis2011,Polimeridis2011b,DEMCEM,Reid2015}, do not suffer from low-frequency inaccuracies and are applicable for both static and dynamic kernels without further modifications.

In the following, we consider the general case of 4-D integrals,
\[
\label{I_s}
I = \int\limits_{E_p}\int\limits_{E_Q} K(\vec r, \vec r') dA_Q dA_P,
\]
where the two quadrilateral (planar or curvilinear) elements $E_P$ and $E_Q$ may coincide (self-term integration), share a common edge (edge adjacent integration), or share a common vertex (vertex adjacent integration). The scalar kernel $K(\vec r, \vec r')$  is typically singular when the observation points $\vec{r}$ coincide with the source points $\vec{r}'$, i.e., $K(\vec r, \vec r') \sim |\vec r- \vec r'|^{-p}$ where $p=1,2$. As evinced by the representative numerical experiments presented in Section VI, the final expressions derived in this work can provide results with very high accuracy. We also show that the overall efficiency can be further improved by a judicious choice of the integration order for each of the four dimensions; the optimal choice of integration orders is left for future work. Finally, in order to enhance reproducibility for computational methods~\cite{Stodden1240}, the complete set of codes used in this paper is available as free, open-source software~\cite{DIRECTFN}.

%% file: formulation.tex
\section{RECTANGULAR PARAMETER SPACE}
\label{sec:square_space}
\begin{figure}[t!]
	\centering
	\subfigure[]{
	\includegraphics[width = 0.47\linewidth]{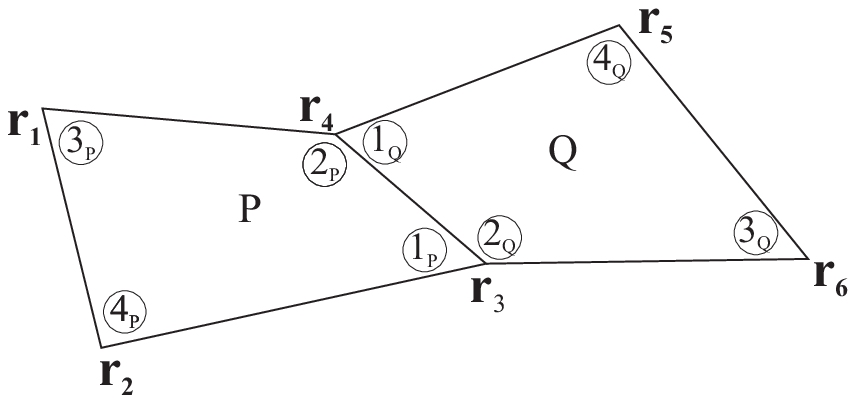}
\label{fig:EA_quad_geom}}
\subfigure[]{
	\includegraphics[width = 0.47\linewidth]{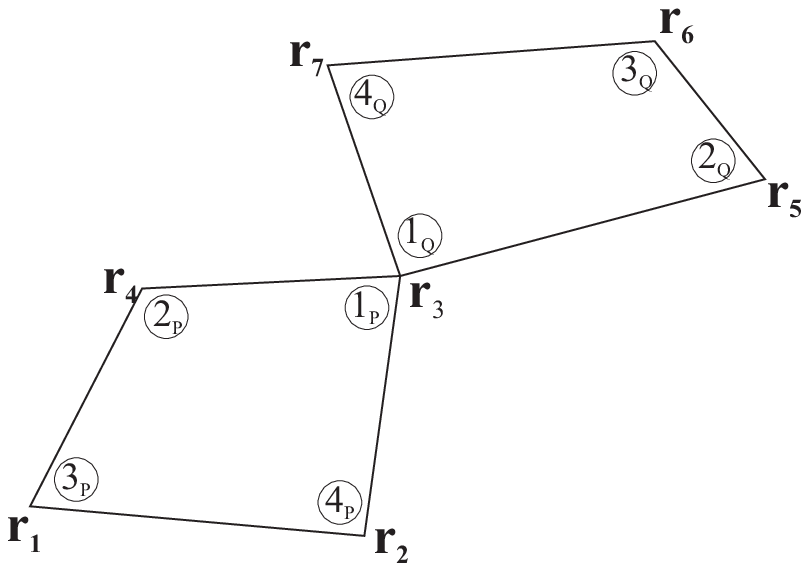}\label{fig:VA_quad_geom}}
	\caption{Orientation of the quadrilateral elements in space: \subref{fig:EA_quad_geom} edge adjacent case; \subref{fig:VA_quad_geom} vertex adjacent case.}
	\label{fig:EA_VA_quad_geom}
\end{figure}

As a first step, we introduce a parametric space $\lbrace u, v \rbrace$, where $-1 \le u \le 1,\quad -1 \le v \le 1$, to transform the original arbitrary quadrilateral to a square. For simplicity we derive all the formulas in this section for planar quadrilaterals, since the extension to curvilinear elements is trivial. The remaining part of the algorithm, described in the next sections, is completely same for both cases.

\[
\vec r(u, v) = \begin{bmatrix}
(1 - u)(1 - v) \vec r_1 + (1 + u)(1 - v) \vec r_2 \\
+ (1+u)(1+v)\vec r_3 +(1-u)(1+v)\vec r_4
\end{bmatrix}/4.
\]
We have to note that this parametrization and all the successive formulas are not only valid for planar elements, but they can also be applied  without any changes for bilinear surfaces~\cite{Kolundzija_book}. The area of the element $dS$ can be expressed as
\[
dS = |\vec r_u \times \vec r_v | du dv,
\]
therefore the associated Jacobian reads
\[
\label{geom_Jacobian}
J(u,v) = |\vec r_u \times \vec r_v |,
\]
where
\{
\label{derivatives}
\vec r_u \equiv \frac{\partial \vec r}{\partial u}= \frac {-\vec r_1 + \vec r_2 + \vec r_3 - \vec r_4 + v\left(\vec r_1 - \vec r_2 + \vec r_3 - \vec r_4\right)}4,\\
\vec r_v \equiv \frac{\partial \vec r}{\partial v} = \frac{-\vec r_1 - \vec r_2 + \vec r_3 + \vec r_4 + u\left(\vec r_1 - \vec r_2 + \vec r_3 - \vec r_4\right)}4.
\}
The original integral (\ref{I_s}) takes the following form in the new parametric space:
\[
\label{I_pq}
I = \int\limits_{-1}^{1} du \int\limits_{-1}^{1} J_P  dv \int\limits_{-1}^{1} du' \int\limits_{-1}^{1} J_Q K(\vec r,\vec r')dv'.
\]
For simplicity, in all successive derivations we will omit the integrands, when no confusion exists. The orientation of the quadrilaterals of the edge adjacent and vertex adjacent cases prior to the square space transformation is shown in Fig.~\ref{fig:EA_VA_quad_geom}.

\section{COINCIDENT INTEGRATION}
\label{sec:ST}
\subsection{First Step}
\begin{figure}[t!]
	\centering
	\includegraphics[width=0.9\linewidth]{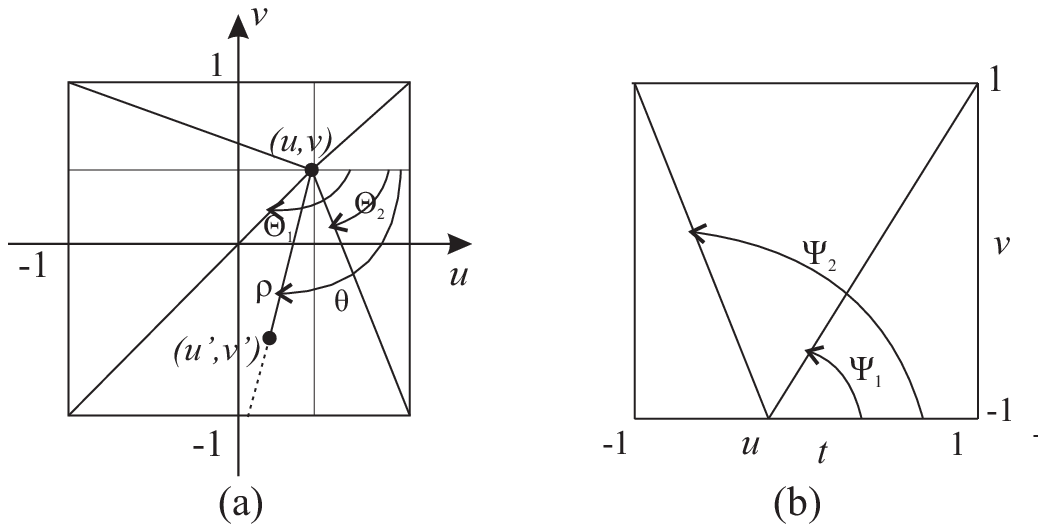}
	\caption{Geometry of the parametric transformations for the coincident case: (a) $\left\lbrace u',v'\right\rbrace \rightarrow \left\lbrace \rho,\theta \right\rbrace$; (b) $\lbrace t,v\rbrace \rightarrow \lbrace \Lambda,\Psi \rbrace$.}
	\label{fig:ST}
\end{figure}

We begin our derivation with introducing a polar coordinate system $\lbrace \rho,\theta \rbrace$ centered at the point $(u,v)$ (depicted schematically in Fig.~\ref{fig:ST}(a)),
\[
u' = u + \rho \cos(\theta), \quad v' = v + \rho \sin(\theta).
\]
Since the upper limit of $\rho$, denoted as $\rho_L$, is different as $\theta$ traverses each edge, the $(\rho,\theta)$ integration must be split in four subtriangles. Here we present only the calculation for the lower subtriangle; the remaining three subtriangles can be handled by rotating the elements accordingly and using the formulas for the lower one, as shown in the following. For the lower subtriangle, the integration limits are 
\[ 0 \le \rho \le \rho_L, \quad \Theta_1 \le \theta \le \Theta_2,\]
where
\[
\begin{split}
\rho_L = \frac{v + 1}{\cos\left(\frac{\pi}{2} + \theta\right)}, \quad \Theta_1 = -\frac{\pi}{2} - \tan^{-1}\left(\frac{u+1}{v+1}\right), \\ \Theta_2 = -\frac{\pi}{2} + \tan^{-1}\left(\frac{1-u}{v+1}\right).
\end{split}
\]
Note that here and below the counter-clockwise angle direction is taken as positive.
Hence, the integral for the lower subtriangle is given by
\[
\label{I_sub_ST}
I^{\rm sub_1} = \int\limits_{-1}^1 du \int\limits_{-1}^1 dv \int\limits_{\Theta_1}^{\Theta_2}  d\theta \int\limits_0^{\rho_L} \rho d\rho.
\]

\subsection{Second Step}

We proceed by introducing the variable t, $-1 \le t \le 1$, via
\[
\begin{aligned}
\theta &= -\frac{\pi}{2} + \tan^{-1}\left(\frac{t-u}{1+v}\right), \\ \frac{d\theta}{dt} &= \frac{1 + v}{(1+v)^2 + (t-u)^2} = F(u,v,t),
\end{aligned}
\]
which results in
\[
\rho_L = \sqrt{(1+v)^2 + (t - u)^2}.
\]
Interchanging the order of integration, (\ref{I_sub_ST}) becomes
\[
\label{I_sub_ST_2}
I^{\rm sub_1} =\int\limits_{-1}^1 du \int\limits_{-1}^1 dt \int\limits_{-1}^1  F(u,v,t) dv\int\limits_0^{\rho_L} \rho d\rho.
\]
Next, a new polar coordinate system $\lbrace \Lambda, \Psi \rbrace$ replaces $\lbrace t, v \rbrace$,
\[
t = u + \Lambda \cos(\Psi), \quad v = -1 + \Lambda\sin(\Psi),
\]
with the Jacobian of this new transformation being $J = \Lambda$. With the two changes of variables, $\theta \rightarrow t$ and $\lbrace t,v\rbrace \rightarrow \lbrace \Lambda, \Psi \rbrace$, we get the following:
\[
\begin{aligned}
\cos(\theta) &\rightarrow \cos(\Psi),\quad \sin(\theta) \rightarrow -\sin(\Psi),\\\rho_L &\rightarrow \Lambda, \quad F \rightarrow \frac{\sin(\Psi)}{\Lambda}.
\end{aligned}
\]

We have to notice that the $\lbrace t,v\rbrace$ domain is a rectangle (Fig.~\ref{fig:ST} (b)), and integrating over $\lbrace\Lambda, \Psi \rbrace$ will necessitate a decomposition into three subdomains: $\Psi_0 \le \Psi \le \Psi_1,\quad \Psi_1 \le \Psi \le \Psi_2,\quad \Psi_2 \le \Psi  \le \Psi_3$, where $\Psi_0 = 0, \quad \Psi_1 = \frac{\pi}{2} - \tan^{-1}\left(\frac{1-u}{2}\right),\quad \Psi_2 = \frac{\pi}{2} + \tan^{-1}\left(\frac{1+u}{2}\right)$ and $\Psi_3 = \pi$, and \eqref{I_sub_ST_2} is written as
\[
I^{\rm sub_1} = \sum\limits_{m=0}^2 \int\limits_{-1}^{1} du \int\limits_{\Psi_m}^{\Psi_{m+1}} \mathcal F(\Psi;\Lambda_{L}) d\Psi ,
\]
where 
\[
\mathcal F(\Psi;\Lambda_{L}) = \sin \Psi \int\limits_0^{\Lambda_L}  d\Lambda \int\limits_0^{\Lambda}  \rho d\rho
\]
is the kernel that is omitted in the following derivations. The limit $\Lambda_L$ for integration over $\Lambda$ depends upon the subdomain (m = 0,1,2) being considered, as shown below.

\subsection{Third step}

The singular integral~(\ref{I_pq}) has been reduced to an integration over $\lbrace u,\Psi\rbrace$ with the $\Psi$ integral decomposed into three subintegrals. The final objective is to regularize further the integral with respect to $u$ by placing it in front of the $\Psi$ integral, so each subintegral has to be examined individually. The subdivision of the integral with respect to $\Psi$ and the limits of integration with respect to $\Lambda$ read
\[
\begin{aligned}
0 \le \Psi \le \Psi_1,&\quad \Lambda_L = \frac{1-u}{\cos(\Psi)},\\
\Psi_1 \le \Psi \le \Psi_2,&\quad \Lambda_L = \frac{2}{\sin{\Psi}},\\
\Psi_2 \le \Psi \le \Psi_3,&\quad \Lambda_L = \frac{1+u}{-\cos{\Psi}}.
\end{aligned}
\]
\subsubsection{Integration over region $0 \le \Psi \le \Psi_1$}
\begin{figure}[t!]
	\centering
	\includegraphics[width=0.9\linewidth]{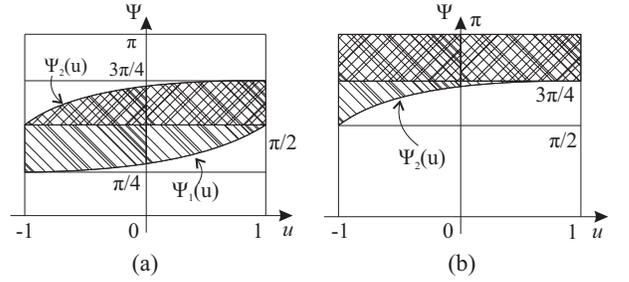}
	\caption{Geometry of the parametric space $\lbrace u,\Psi\rbrace$: (a) $0 \le \Psi \le \Psi_1$ and $\Psi_1 \le \Psi \le \Psi_2$; (b) $\Psi_2 \le \Psi \le \pi$.}
	\label{fig:Psi_u_ST}
\end{figure}

The domain of integration is depicted schematically in Fig.~\ref{fig:Psi_u_ST}, below the curve $\Psi_1(u)$. After interchanging the $u$ and $\Psi$ integration, we obtain
\begin{equation}
\int\limits_{-1}^1 du \int\limits_0^{\Psi_1} d\Psi  = \int\limits_0^{\frac{\pi}{4}}d\Psi\int\limits_{-1}^1  du + \int\limits_{\frac{\pi}{4}}^{\frac{\pi}{2}}d\Psi\int\limits_{{u_1}_{\psi}}^1  du,
\end{equation}
where
\[
{u_1}_{\psi} = 2\tan\left(\Psi - \frac{\pi}{2}\right) + 1.
\]
\subsubsection{Integration over region $\Psi_1 \le \Psi \le \Psi_2$}
The domain of integration is shown again in Fig.~\ref{fig:Psi_u_ST}(a). After re-ordering the integration, we get the following two integrals:
\begin{equation}
\int\limits_{-1}^1 du \int\limits_{\Psi_1}^{\Psi_2} d\Psi =
	\int\limits_{\frac{\pi}{4}}^{\frac{\pi}{2}}d\Psi\int\limits_{-1}^{{u_1}_{\psi}}  du + \int\limits_{\frac{\pi}{2}}^{\frac{3\pi}{4}}d\Psi\int\limits_{{u_2}_{\psi}}^1  du,
\end{equation}
where
\[
{u_1}_{\psi} = 2\tan\left(\Psi - \frac{\pi}{2}\right) + 1,\,
{u_2}_{\psi} = 2\tan\left(\Psi - \frac{\pi}{2}\right) - 1.
\]

\subsubsection{Integration over region $\Psi_2 \le \Psi \le \pi$}
The domain of integration in this case is depicted in Fig.~\ref{fig:Psi_u_ST}(b). After re-ordering the integration, we obtain
\begin{equation}
\int\limits_{-1}^1 du \int\limits_{\Psi_2}^{\pi}  d\Psi =
\int\limits_{\frac{\pi}{2}}^{\frac{3\pi}{4}}d\Psi\int\limits_{-1}^{{u_2}_{\psi}}  du + \int\limits_{\frac{3\pi}{4}}^{\pi}d\Psi\int\limits_{-1}^1 du,
\end{equation}
where
\[
{u_2}_{\psi} = 2\tan\left(\Psi - \frac{\pi}{2}\right) - 1.
\]

\subsection{Final Formulas}

Finally, the singular integral~(\ref{I_pq}) for the lower subtriangle has been reduced to the following 6 sufficiently smooth integrals:
\[
\begin{aligned}
I^{\rm sub_1} &= \int\limits_0^{\frac{\pi}{4}}d\Psi\int\limits_{-1}^1  du + \int\limits_{\frac{\pi}{4}}^{\frac{\pi}{2}}d\Psi\int\limits_{{u_1}_{\psi}}^1 du 
+ \int\limits_{\frac{\pi}{4}}^{\frac{\pi}{2}}d\Psi\int\limits_{-1}^{{u_1}_{\psi}}  du \\
&+ \int\limits_{\frac{\pi}{2}}^{\frac{3\pi}{4}}d\Psi\int\limits_{{u_2}_{\psi}}^1  du 
+ \int\limits_{\frac{\pi}{2}}^{\frac{3\pi}{4}}d\Psi\int\limits_{-1}^{{u_2}_{\psi}}  du + \int\limits_{\frac{3\pi}{4}}^{\pi}d\Psi\int\limits_{-1}^1  du,
\end{aligned}
\]
where 
\[
{u_1}_{\psi} = 2\tan\left(\Psi - \frac{\pi}{2}\right) + 1,\,
{u_2}_{\psi} = 2\tan\left(\Psi - \frac{\pi}{2}\right) - 1.
\]
The overall Jacobian after all parametric transformations is given by
\[
\mathcal{J}^{\rm ST} = (J_P J_Q)\rho \sin \Psi,
\]
while the original variables take the following form:
\[
\begin{aligned}
u &\rightarrow u,\, v \rightarrow \Lambda\sin(\Psi) - 1,\\
u' &\rightarrow u + \rho\cos(\Psi),\, v' = -\rho\sin(\Psi) + \Lambda\sin(\Psi) - 1.
\end{aligned}
\]
Exploiting the symmetry of the rectangular parameter space, we can derive the formulas for the other three subtriangles by simply rotating them accordingly and employing the formulas for the lower one. Hence, the final formula for the original singular integral \eqref{I_pq} is given by
\[
I^{\rm ST} = I^{\rm sub_1} + I^{\rm sub_2} + I^{\rm sub_3} + I^{\rm sub_4},
\]
where
\[
I^{\rm sub_2} = \left.I^{\rm sub_1}\right|_{\begin{bmatrix}u\\v\end{bmatrix} \rightarrow\begin{bmatrix}
0 & -1 \\ 1 & 0\\
\end{bmatrix}\begin{bmatrix}u\\v\end{bmatrix} },
\]
\[
I^{\rm sub_3} = \left.I^{\rm sub_1}\right|_{\begin{bmatrix}u\\v\end{bmatrix} \rightarrow\begin{bmatrix}
	0 & -1 \\ -1 & 0\\
	\end{bmatrix}\begin{bmatrix}u\\v\end{bmatrix} }
\]
and
\[
I^{\rm sub_4} = \left.I^{\rm sub_1}\right|_{\begin{bmatrix}u\\v\end{bmatrix} \rightarrow\begin{bmatrix}
	0 & 1 \\ -1 & 0\\
	\end{bmatrix}\begin{bmatrix}u\\v\end{bmatrix} }.
\]

\section{EDGE ADJACENT INTEGRATION}
\label{sec:EA}
\subsection{First Step}
Based upon the coincident integration scheme, we employ a polar coordinate transformation for the inner integration to cancel the line of singularity defined by $v = v' = -1$ and $u = -u'$,
\[
u' = \rho\cos(\theta) - u,\quad v' = \rho\sin(\theta) - 1.
\]
The integration with respect to $\theta$ should be split into three terms, as illustrated in Fig.~\ref{fig:EA}(a):
\[
I = I_{\rm sub_1} + I_{\rm sub_2} + I_{\rm sub_3},
\]
where
\[
\begin{aligned}
I_{\rm sub_1} &= \int\limits_{-1}^{1}du\int\limits_{-1}^{1} dv\int\limits_{0}^{\Theta_1(u)}d\theta\int\limits_{0}^{L_1}\rho d\rho,\\
I_{\rm sub_2} &= \int\limits_{-1}^{1} du \int\limits_{-1}^{1}dv\int\limits_{\Theta_1(u)}^{\Theta_2(u)}d\theta\int\limits_{0}^{L_2}\rho d\rho,\\
I_{\rm sub_3} &= \int\limits_{-1}^{1}du\int\limits_{-1}^{1}dv\int\limits_{\Theta_2(u)}^{\pi}d\theta\int\limits_0^{L_3}\rho d\rho
\end{aligned}
\]
and
\[
\begin{aligned}
\Theta_1(u) &= \frac{\pi}{2} - \tan^{-1}\left(\frac{1 + u}{2}\right),\\
\Theta_2(u) &= \frac{\pi}{2} + \tan^{-1}\left(\frac {1-u}2\right),\\
\end{aligned}
\]
\[
L_1 = \frac{1+u}{\cos(\theta)},\quad
L_2 = \frac 2 {\sin(\theta)},\quad
L_3 = \frac {u - 1}{\cos(\theta)}.
\]

Since the break-points in $\theta$ are only functions of $u$, the integration can be rearranged as follows:
\[
\begin{aligned}
I_{\rm sub_1} &= \int\limits_{-1}^{1}du\int\limits_0^{\Theta_1(u)}d\theta\int\limits_{-1}^{1}dv\int\limits_{0}^{L_1}\rho d\rho,\\
I_{\rm sub_2} &= \int\limits_{-1}^{1}du\int\limits_{\Theta_1(u)}^{\Theta_2(u)}d\theta\int\limits_{-1}^{1}dv\int\limits_{0}^{L_2}\rho d\rho,\\
I_{\rm sub_3} &= \int\limits_{-1}^{1}du\int\limits_{\Theta_2(u)}^{\pi}d\theta\int\limits_{-1}^{1}dv\int\limits_0^{L_3}\rho d\rho.\\
\end{aligned}
\]
As the singularity now occurs when $v = -1$ and $\rho = 0$, we proceed by introducing a second polar coordinate transformation,
\[
\rho = \Lambda\cos(\Psi),\quad v = -1 + \Lambda\sin(\Psi),\quad J_2 = \Lambda.
\]
The original integral can be written as a sum:
\[
\label{2sum}
I = \sum\limits_{l=0}^2\sum\limits_{m=0}^1\int\limits_{-1}^1 du \int\limits_{\Theta_l}^{\Theta_{l+1}} d\theta \int\limits_{\Psi_m}^{\Psi_{m+1}} {\mathcal G(\Psi;\Lambda_{L})} d\Psi,
\]
where
\begin{equation}
\label{mathcalG}
\mathcal G(\Psi;\Lambda_{L}) = \cos \Psi \int\limits_{0}^{\Lambda_L} \Lambda^2 d\Lambda
\end{equation}
can be evaluated numerically, and is omitted in the following derivations. The integration limits in \eqref{2sum} are given by
\[
\begin{aligned}
\Theta_0 &= 0,\quad \Theta_1 =\frac{\pi}{2} - \tan^{-1}\frac{1 + u}{2},\\ \Theta_2 &= \frac{\pi}{2} + \tan^{-1}\frac {1-u}2,\quad \Theta_3 = \pi,  
\end{aligned}
\]
and
\[
\Psi_0 = 0,\quad \Psi_1 = \tan^{-1}\left(\frac{2}{L(u,\theta)}\right),\quad \Psi_2 = \frac {\pi}2,
\]
while the integration limits with respect to $\Lambda$ are given below.

\subsection{Second Step}

\subsubsection{Integration over region $\Theta_1 \le\theta \le \Theta_2$}
\begin{figure}[t!]
	\includegraphics[width=0.9\linewidth]{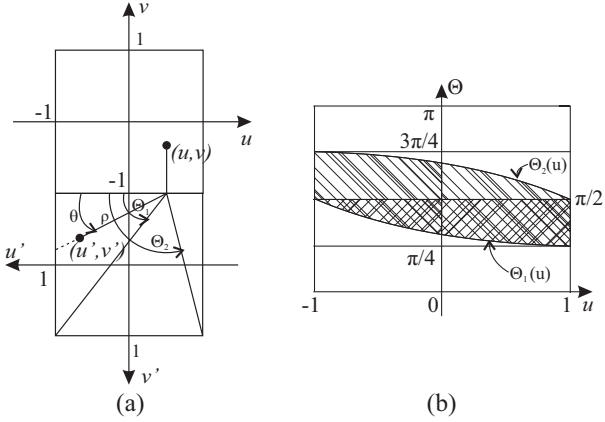}
	\caption{Edge adjacent integration: (a) polar coordinate transformation $\left\lbrace u',v'\right\rbrace \rightarrow \left\lbrace \rho,\theta \right\rbrace$; (b) the $\left\lbrace u, \theta \right\rbrace$ domain for the second shift of the integral, $\Theta_1 \le \theta \le \Theta_2$.}
	\label{fig:EA}
\end{figure}

In this case, the splitting of $\Psi$ integrals is independent of $u$ and the integral is given by
\begin{equation}
I^{\theta_{12}} = 
\int\limits_{-1}^1 du \int\limits_{\Theta_1}^{\Theta_2} d\theta \int\limits_0^{\Psi_1^{12}} d\Psi 
+ \int\limits_{-1}^1 du \int\limits_{\Theta_1}^{\Theta_2} d\theta\int\limits_{\Psi_1^{12}}^{\frac{\pi}2}  d\Psi,
\end{equation}
where the upper limit of $\Lambda$ in \eqref{mathcalG} is different in the two terms:
\[
\Lambda_L = \begin{cases}
\frac{L_2}{\cos(\Psi)} = \frac{2}{\sin(\theta)\cos(\Psi)},&\quad 0 < \Psi < \Psi_1^{12},\\
\frac{2}{\sin{\Psi}},& \quad \Psi_1^{12} < \Psi < \frac{\pi}2.
\end{cases}
\]
Moreover,
\{
\label{theta_u}
{\theta_1}_u \equiv \Theta_1 = \frac{\pi}{2} - \tan^{-1}\left(\frac{1 + u}{2}\right),\\
{\theta_2}_u \equiv \Theta_2 = \frac{\pi}{2} + \tan^{-1}\left(\frac {1-u}2\right),
\}
and the integral with respect to $\Psi$ is split at
\[
\Psi_1^{12} = \tan^{-1}\left(\frac{2}{L_2}\right) = \tan^{-1}\left(\sin\theta\right).
 \]
Hence, once $u$ and $\theta$ are interchanged, the $u$ can be moved immediately past the $\Psi$ integral. Noting that ${\theta_1}_u(-1) =  \frac{\pi}{2},\quad {\theta_1}_u(1) = \frac{\pi}{4},\quad {\theta_2}_u(-1) = \frac{3\pi}{4},\quad {\theta_2}_u(1) = \frac{\pi}{2},$
the geometry for interchanging $u$ and $\theta$ is shown in Fig.~\ref{fig:EA}(b). Inverting the relationships between $u$ and $\theta$ yields
\[
\label{u_theta}
{u_1}_{\theta} = 2\tan\left(\frac{\pi}{2} - \theta\right) - 1, \quad {u_2}_{\theta} =2\tan\left(\frac{\pi}{2} - \theta\right) + 1,
\]
while switching the integrals results in
\begin{multline}
I^{\theta_{12}} = 
\underbrace{\int\limits_{\frac{\pi}{4}}^{\frac{\pi}{2}} d\theta \int\limits_0^{\Psi_1^{12}} \!d\Psi \int\limits_{{u_1}_{\theta}}^1 du + \int\limits_{\frac{\pi}{2}}^{\frac{3\pi}{4}}d\theta \int\limits_0^{\Psi_1^{12}} d\Psi \int\limits_{-1}^{{u_2}_{\theta}}  du}_{I^{\theta_{12},\Psi^-}} \\
+ \underbrace{\int\limits_{\frac{\pi}{4}}^{\frac{\pi}{2}} d\theta \int\limits_{\Psi_1^{12}}^{\frac{\pi}{2}} d\Psi \int\limits_{{u_1}_{\theta}}^1  du + \int\limits_{\frac{\pi}{2}}^{\frac{3\pi}{4}}d\theta \int\limits_{\Psi_1^{12}}^{\frac{\pi}{2}} d\Psi \int\limits_{-1}^{{u_2}_{\theta}}  du}_{I^{\theta_{12},\Psi^+}}. 
\end{multline}
\subsubsection{Integration over region $0 < \theta \le \Theta_1$}
In this case, the breakpoint in $\Psi$ is a function of $\theta$ and $u$ both, and re-ordering of integrations will produce eight integrals. The two first integrals are given by
\begin{equation}
I^{\theta_1}  =
 \int\limits_{-1}^1 du \int\limits_0^{\Theta_1} d\theta \int\limits_0^{\Psi_1^1}  d\Psi 
 + \int\limits_{-1}^1 du \int\limits_{0}^{\Theta_1} d\theta\int\limits_{\Psi_1^{1}}^{\frac{\pi}2} d\Psi,
\end{equation}
where
\[
\Lambda_L = \begin{cases}
\frac{L_1}{\cos(\Psi)} = \frac{1+u}{\cos(\theta)\cos(\Psi)},&\quad 0 < \Psi < \Psi_1^1,\\
\frac{2}{\sin(\Psi)}, & \quad \Psi_1^1 < \Psi < \frac{\pi}{2},\\
\end{cases}
\]
and
\[
\Psi_1^1  = \frac{\pi}2 - \tan^{-1}\left(\frac{1+u}{2\cos(\theta)}\right).
\]
As in the previous section, the $\theta$ and $u$ integrals are easily interchanged. The domain of integration is depicted schematically at Fig.~\ref{fig:EA}(b), below the curve $\Theta_1(u)$. This results in the following four integrals:
\begin{multline}
\label{I_theta1}
I^{\theta_1} = \underbrace{\int\limits_0^{\frac{\pi}{4}} d\theta\int\limits_{-1}^1 du
\int\limits_0^{\Psi_1^1}   d\Psi + \int\limits_{\frac{\pi}{4}}^{\frac{\pi}{2}} d\theta\int\limits_{-1}^{{u_1}_{\theta}} du
\int\limits_0^{\Psi_1^1}  d\Psi}_{I^{\theta_1,\Psi^-}} \\
+ \underbrace{\int\limits_0^{\frac{\pi}{4}} d\theta\int\limits_{-1}^1 du\int\limits_{\Psi_1^1}^{\frac{\pi}{2}}  d\Psi + \int\limits_{\frac{\pi}{4}}^{\frac{\pi}{2}} d\theta\int\limits_{-1}^{{u_1}_{\theta}} du
\int\limits_{\Psi_1^1}^{\frac{\pi}{2}}  d\Psi}_{I^{\theta_1,\Psi^+}},
\end{multline}
where the expression for ${u_1}_{\theta}$ is given in \eqref{u_theta}. The next step is to regularize further the integral by interchanging of $u$ and $\Psi$.

\begin{figure}[t!]
	\centering
	\includegraphics[width=0.9\linewidth]{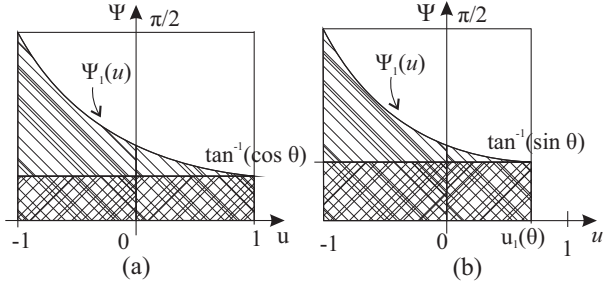}
	\caption{Polar coordinate transformations employed in the edge adjacent integration: the domain for interchanging the integrals $\left\lbrace u,\Psi\right\rbrace$, for a fixed value of $\theta$ ($0 
		\le \theta \le \Theta_1$): (a) $u_1(\theta) = 1$ ($0 \le \theta \le \pi/4$); (b) $u_1(\theta) < 1$ ($\pi/4 < \theta \le \pi/2$).}
	\label{fig:Psi_u}
\end{figure}

\paragraph{Integration over region $0 < \Psi \le \Psi_1^1$} The domain under consideration is shown in Figs.~\ref{fig:Psi_u}(a-b), below the curve $\Psi_1(u)$. 
Moving the $u$ integral to the front in the first two integrals in \eqref{I_theta1}, corresponding to the case of $0 < \Psi \le \Psi_1^1$, results in
\begin{multline}
I^{\theta_1,\Psi^-}
 = 
 \int\limits_0^{\frac{\pi}{4}} d\theta \int\limits_{0}^{\Psi_{\theta}^1} d\Psi \int\limits_{-1}^1  du +
\int\limits_0^{\frac{\pi}{4}} d\theta \int\limits_{\Psi_{\theta}^1}^{\frac{\pi}{2}} d\Psi \int\limits_{-1}^{{u_1}_{\psi}}  du \\
+ \int\limits_{\frac{\pi}{4}}^{\frac{\pi}{2}} d\theta \int\limits_{0}^{\Psi_{\theta}^2} d\Psi \int\limits_{-1}^{{u_1}_{\theta}}  du +
\int\limits_{\frac{\pi}{4}}^{\frac{\pi}{2}} d\theta \int\limits_{\Psi_{\theta}^2}^{\frac{\pi}{2}} d\Psi \int\limits_{-1}^{{u_1}_{\psi}} du,
\end{multline}
where
\[
\begin{aligned}
{u_1}_{\psi} &= 2\cos(\theta)\cdot\tan\left(\frac{\pi}{2} - \Psi\right) - 1,\\
{u_1}_{\theta} &= 2\tan\left(\frac{\pi}{2} - \theta\right) - 1,\\
\Psi_{\theta}^1 &= \left.\Psi_1^1\right|_{u = 1} = \tan^{-1}(\cos(\theta)),\\
\Psi_{\theta}^2 &= \left.\Psi_1^1\right|_{u = {u_1}_{\theta}} = \tan^{-1}(\sin(\theta)).
\end{aligned}
\]
\paragraph{Integration over region $\Psi_1^1 < \Psi \le \pi/2 $} After interchanging $u$ and $\Psi$ the last two integrals in \eqref{I_theta1}, corresponding to $\Psi_1^1 < \Psi \le \pi/2$ (the region depicted in Figs.~\ref{fig:Psi_u}(a-b) under the curve $\Psi_1(u)$), become

\begin{equation}
I^{\theta_1,\Psi^+} =
 \int\limits_0^{\frac{\pi}{4}} d\theta \int\limits_{\Psi_{\theta}^1}^{\frac{\pi}{2}} d\Psi \int\limits_{{u_1}_{\psi}}^1 du + 
\int\limits_{\frac{\pi}{4}}^{\frac{\pi}{2}} d\theta \int\limits_{\Psi_{\theta}^2}^{\frac{\pi}{2}} d\Psi \int\limits_{{u_1}_{\psi}}^{{u_1}_{\theta}} du.
\end{equation}

\subsubsection{Integration over region $\Theta_2 < \theta \le \pi$}

\begin{figure}[t!]
	\centering
	\includegraphics[width=0.9\linewidth]{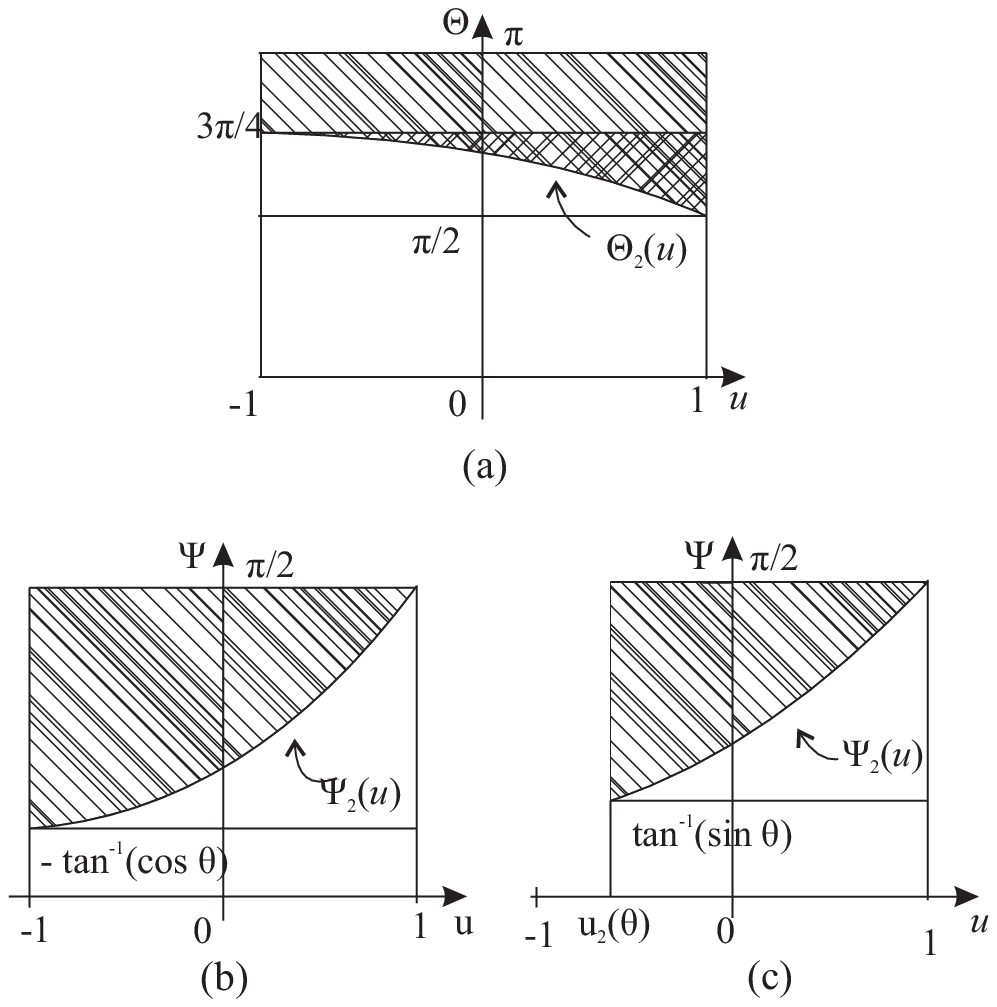}
	\caption{Polar coordinate transformations employed in the edge adjacent integration: (a) the $\left\lbrace u, \theta \right\rbrace$ domain for the third shift of the integral, $\Theta_2 \le \theta \le \pi$; (b) the domain for interchanging the integrals $\left\lbrace u,\Psi\right\rbrace$, for a fixed value of $\theta$ and $u_2(\theta) = -1$ ($\pi/2 \le \theta \le 3\pi/4$); and (c) the domain for interchanging the integrals $\left\lbrace u,\Psi\right\rbrace$, for a fixed value of $\theta$ and $u_2(\theta) > -1$ ($3\pi/4 < \theta \le \pi$).}
	\label{fig:Psi_u_2_EA}
\end{figure}

This case is similar to the previous one. The two first integrals are given by 
\begin{equation}
I^{\theta_2} =
 \int\limits_{-1}^1 du \int\limits_{\Theta_2}^{\pi} d\theta \int\limits_0^{\Psi_1^2}  d\Psi 
+ \int\limits_{-1}^1 du \int\limits_{\Theta_2}^{\pi} d\theta \int\limits_{\Psi_1^2}^{\frac{\pi}{2}}  d\Psi,
\end{equation}
where the upper limit of $\Lambda$ is
\[
\Lambda_L = \begin{cases}
\frac{L_3}{\cos(\Psi)} = \frac{u-1}{\cos(\theta)\cos(\Psi)},& \quad 0 < \Psi < \Psi_1^2,\\
\frac{2}{\sin(\Psi)},& \quad \Psi_1^2 < \Psi < \frac{\pi}{2},
\end{cases}
\]
and
\[
\Psi_1^2 = \frac{\pi}{2} - \tan^{-1}\left(\frac{u-1}{2\cos(\theta)}\right).
\]
The $\theta$ and $u$ integrals can be interchanged, and the domain is depicted in Fig.~\ref{fig:Psi_u_2_EA}(a) under the curve $\Theta_2(u)$. 

After the interchanging we obtain
\begin{multline}
\label{I_theta2}
I^{\theta_2}  =\underbrace{\int\limits_{\frac{\pi}{2}}^{\frac{3\pi}{4}} d\theta \int\limits_{{u_2}_{\theta}}^1 du \int\limits_0^{\Psi_1^2}d\Psi + \int\limits_{\frac{3\pi}{4}}^{\pi} d\theta \int\limits_{-1}^1 du \int\limits_0^{\Psi_1^2}  d\Psi}_{I^{\theta_2,\Psi^-}}\\
+ \underbrace{\int\limits_{\frac{\pi}{2}}^{\frac{3\pi}{4}} d\theta \int\limits_{{u_2}_{\theta}}^1 du \int\limits_{\Psi_1^2}^{\frac{\pi}{2}}  d\Psi + \int\limits_{\frac{3\pi}{4}}^{\pi} d\theta \int\limits_{-1}^1 du \int\limits_{\Psi_1^2}^{\frac{\pi}{2}}  d\Psi}_{I^{\theta_2,\Psi^+}}, 
\end{multline}
where ${u_2}_{\theta}$ is given in \eqref{u_theta}. The final step is the interchanging of $u$ and $\Psi$, and the associated geometry is shown in Figs.~\ref{fig:Psi_u_2_EA}(b-c).

\paragraph{Integration over region $0 < \Psi \le \Psi_1^2$}
After moving the $u$ integral to the front, the first two integrals in \eqref{I_theta2} read
\begin{multline}
I^{\theta_2,\Psi^-} =  \int\limits_{\frac{\pi}{2}}^{\frac{3\pi}{4}} d\theta \int\limits_0^{\Psi_{\theta}^2} d\Psi \int\limits_{{u_2}_{\theta}}^1  du + \int\limits_{\frac{\pi}{2}}^{\frac{3\pi}{4}} d\theta \int\limits_{\Psi_{\theta}^2}^{\frac{\pi}{2}} d\Psi \int\limits_{{u_2}_{\psi}}^1 du \\
+ \int\limits_{\frac{3\pi}{4}}^{\pi} d\theta \int\limits_0^{-\Psi_{\theta}^1} d\Psi \int\limits_{-1}^1  du +
\int\limits_{\frac{3\pi}{4}}^{\pi} d\theta \int\limits_{-\Psi_{\theta}^1}^{\frac{\pi}{2}} d\Psi \int\limits_{{u_2}_{\psi}}^1 du,
\end{multline}
where
\[
\begin{aligned}
{u_2}_{\psi} &= 2\cos(\theta)\cdot \tan\left(\frac{\pi}{2} - \Psi\right) + 1,\\
{u_2}_{\theta} &= 2\tan\left(\frac{\pi}{2} - \theta\right) + 1,\\
-\Psi_{\theta}^1 &= \left.\Psi_1^2\right|_{u = -1}  = -\tan^{-1}(\cos(\theta)),\\
\Psi_{\theta}^2 &= \left.\Psi_1^2\right|_{u = {u_2}_{\theta}}  = \tan^{-1}(\sin(\theta)).
\end{aligned}
\]

\paragraph{Integration over region $\Psi_1^2 < \Psi \le \pi/2$} 
The last two integrals in \eqref{I_theta2}, corresponding to the case of $\Psi_1^2 < \Psi \le \pi/2$, become
\begin{equation}
I^{\theta_2,\Psi^+} = 
\int\limits_{\frac{\pi}{2}}^{\frac{3\pi}{4}} d\theta \int\limits_{\Psi_{\theta}^2}^{\frac{\pi}{2}} d\Psi \int\limits_{{u_2}_{\theta}}^{{u_2}_{\psi}} du
+ \int\limits_{\frac{3\pi}{4}}^{\pi} d\theta \int\limits_{-\Psi_{\theta}^1}^{\frac{\pi}{2}} d\Psi \int\limits_{-1}^{{u_2}_{\psi}} du.
\end{equation}
The combined Jacobian in the edge-adjacent case reads
\[
\mathcal I^{\rm EA} = (J_P J_Q) \Lambda^2 \cos \Psi 
\]
and the original variables can be written as
\[
\begin{aligned}
u &\rightarrow u, \quad v \rightarrow -1 + \Lambda \sin \Psi,\\
u' &\rightarrow \Lambda \cos \Psi \cos \theta - u, \quad v' \rightarrow \Lambda \cos \Psi \sin \theta - 1.
\end{aligned}
\]
\section{VERTEX ADJACENT INTEGRATION}
\label{sec:VA}
\begin{figure}[t!]
	\centering
	\includegraphics[width = 0.7\linewidth]{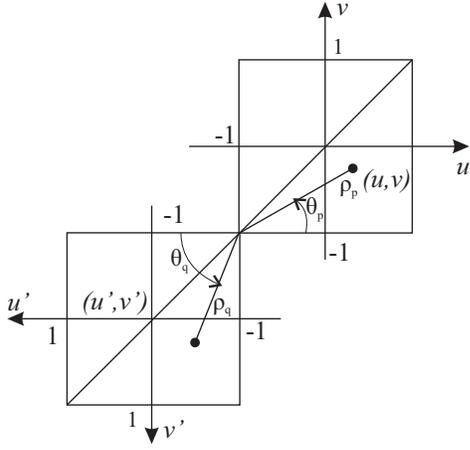}
	\caption{Polar coordinate transformations employed in vertex adjacent integration: $\lbrace u,v\rbrace \rightarrow \lbrace \rho_p,\theta_p\rbrace,\lbrace u',v'\rbrace\rightarrow \lbrace\rho_q,\theta_q\rbrace$.}
	\label{fig:VA}
\end{figure}

In the case where the source and observation quadrilaterals share only a single vertex, we begin by orienting the elements so that the singular point is at $u = u' = -1,\quad v = v' = -1$. Next, we introduce a separate coordinate system for each element, as shown at Fig.~\ref{fig:VA}:
\[
\begin{aligned}
u &= -1 + \rho_p \cos(\theta_p),\quad v = -1 + \rho_p \sin(\theta_p), \\
u' &= -1 + \rho_q \cos(\theta_q),\quad v' = -1 + \rho_q\sin(\theta_q).
\end{aligned}
\]
This results in four integrals:
\[
\begin{aligned}
I &= \int\limits_0^{\frac{\pi}{4}} d\theta_p \int\limits_0^{L_p^1} \rho_p d\rho_p \int\limits_0^{\frac{\pi}{4}} d\theta_q \int\limits_0^{L_q^1} \rho_q d\rho_q \\&+ \int\limits_0^{\frac{\pi}{4}} d\theta_p \int\limits_0^{L_p^1} \rho_p d\rho_p\int\limits_{\frac{\pi}{4}}^{\frac{\pi}{2}} d\theta_q \int\limits_0^{L_q^2}  \rho_q d\rho_q \\&+ 
 \int\limits_{\frac{\pi}{4}}^{\frac{\pi}{2}} d\theta_p \int\limits_0^{L_p^2}  \rho_p d\rho_p  \int\limits_0^{\frac{\pi}{4}} d\theta_q \int\limits_0^{L_q^1} \rho_q d\rho_q \\&+  \int\limits_{\frac{\pi}{4}}^{\frac{\pi}{2}} d\theta_p \int\limits_0^{L_p^2}  \rho_p d\rho_p \int\limits_{\frac{\pi}{4}}^{\frac{\pi}{2}} d\theta_q \int\limits_0^{L_q^2}  \rho_q d\rho_q,
\end{aligned}
\]
where
\[
\begin{split}
L_p^1 = \frac{2}{\cos(\theta_p)} ,\quad L_p^2 = \frac{2}{\sin(\theta_p)},\\
L_q^1 = \frac{2}{\cos(\theta_q)}, \quad L_q^2 = \frac{2}{\sin(\theta_q)}.
\end{split}
\]
The singularity is at the common vertex $\rho_p = \rho_q = 0$, so it's reasonable to use a polar coordinate transformation,
\[
\rho_p = \Lambda\cos(\Psi),\quad \rho_q = \Lambda\sin(\Psi).
\]
Since the $\lbrace\rho_p, \rho_q\rbrace$ domain is rectangular, the $\Psi$ integration must be split into two pieces, which leads to the final eight integrals,
\begin{multline}
I = \sum\limits_{m=1}^2\sum\limits_{n=1}^2
\int\limits_{\Theta_{m-1}}^{\Theta_m} d\theta_p\int\limits_{\Theta_{n-1}}^{\Theta_n} d\theta_q \\
\times\left[ \int\limits_0^{\Psi_1^{m,n}} \mathcal H(\Psi;L_1^{m,n})  d\Psi  
+  \int\limits_{\Psi_1^{m,n}}^{\frac{\pi}{2}} \mathcal H(\Psi;L_2^{m,n}) d\Psi \right],
\end{multline}
where
\[
\mathcal H(\Psi;L_i^{m,n}) = \cos \Psi \sin \Psi \int\limits_0^{L_i^{m,n}(\Psi)} \Lambda^3 d\Lambda, \quad i = 1,2,
\]
and the integration limits are given by
\[
\begin{aligned}
&\Theta_0 = 0,\quad \Theta_1 = \frac{\pi}{4},\quad \Theta_2 = \frac{\pi}{2},\\
&L_1^{m,n}(\Psi) = \frac{L_p^m(\theta_p)}{\cos(\Psi)},\quad
L_2^{m,n}(\Psi) = &\frac{L_q^n(\theta_q)}{\sin(\Psi)},\\
&\Psi_1^{m,n} = \tan^{-1}\left(\frac{L_q^{n}}{L_p^{m}}\right).
\end{aligned}
\]
The final Jacobian takes the form
\[
\mathcal J^{\rm VA} = (J_P J_Q) \Lambda^3 \cos \Psi \sin \Psi 
\]
and the original variables are given by
\[
\begin{aligned}
u &\rightarrow -1 + \Lambda \cos \Psi \cos \theta_p, \, v \rightarrow -1 + \Lambda \cos \Psi \sin \theta_p,\\
u' &\rightarrow -1 + \Lambda \sin \Psi \cos \theta_q, \, v' \rightarrow -1 + \Lambda \sin \Psi \sin \theta_q.
\end{aligned}
\]

%% file: results.tex
\section{NUMERICAL RESULTS}
\label{sec:results}
This section presents various examples that demonstrate
the convergence properties and computational efficiency of the proposed algorithms for
both weakly singular and strongly singular integrals arising in
Galerkin SIE formulations.

\subsection{Comparison with DIRECTFN for triangles}

\begin{figure}[t!]
 	\includegraphics[width = 1.0\linewidth]{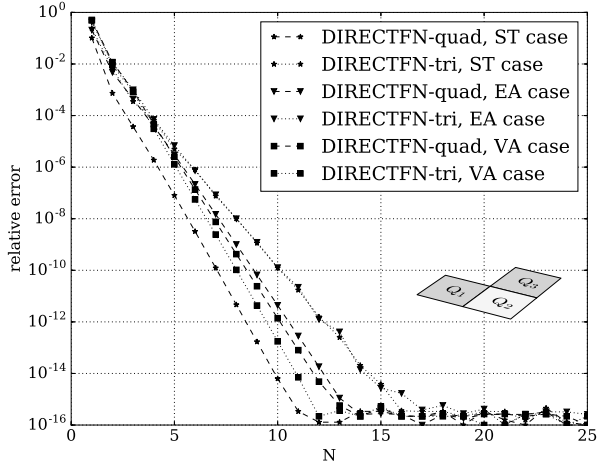}
 	\caption{Relative error in computing the singular integrals \eqref{I_WS} as a function of the order of the 1-D Gaussian quadrature rules, DIRECTFN-quad comparison with DIRECTFN-tri.}
 	\label{fig:Error_all}
 \end{figure}

In the first set of numerical experiments we demonstrate the effectiveness of the proposed algorithm in terms of convergence rate and the computational efficiency. More specifically, we compute the following weakly singular integral:
\[
\label{I_WS}
I^{\rm WS} = \int\limits_{E_P}\int\limits_{E_Q} G(\vec r, \vec r') dS' dS,
\] 
where $G(\vec r, \vec r') = \frac{e^{-ik|\vec r - \vec r'|}}{4\pi |\vec r - \vec r'|}$ is the free-space Green function and $E_P$ and $E_Q$ are observation and source quadrilateral elements. As reference, we use the results obtained by the  method presented herein (dubbed DIRECTFN-quad) with a high order of Gaussian quadrature for all four one-dimensional integrations, i.e., $N_1=N_2=N_3=N_4=25$. We compare the convergence of the novel algorithms with the original DIRECTFN method (dubbed DIRECTFN-tri)~\cite{DIRECTFN}, applied to the combination of the triangles obtained by splitting accordingly the quadrilaterals $E_P$ and $E_Q$. The singular integral~\eqref{I_WS} is computed for all possible configurations, i.e. $E_P \equiv Q_1$ and $E_Q \equiv Q_1,Q_2,Q_3$ for ST, EA and VA elements, respectively. All squares $Q_i$ have sides with length $d = 0.1\lambda$, where $\lambda$ is the wavelength associated to the operating frequency. The relative errors, defined as
\[
\label{max_err}
\varepsilon = \left\|\frac{\|I - I_{\mathrm{ ref}}\|_2}{\|I_{\mathrm{ref}}\|_2} + \epsilon\right\|_2,
\]
(with $\|. \|_2$ being the 2-norm and $\epsilon$ the machine epsilon) are presented in Fig.~\ref{fig:Error_all}, where one can clearly observe the exponential convergence with respect to the integration order. This behavior suggests that the kernels are sufficiently smooth after the series of transformations and the reordering of the integrations. 

\subsection{Weakly and strongly singular integrals with vector basis functions}
\begin{figure}[t!]
	\includegraphics[width = 1.0\linewidth]{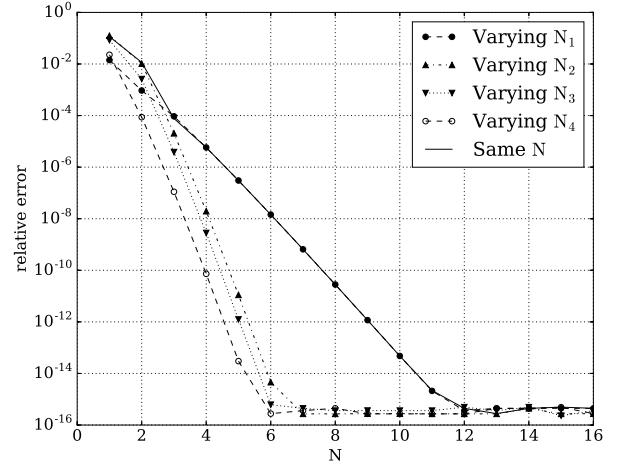}
	\caption{Relative error in computing the weakly singular integrals \eqref{I_mn_WS} over coincident squares as a function of the order of the 1-D Gaussian quadrature.}
	\label{fig:error_WS_one_ST}
\end{figure}

\begin{figure}[t!]
	\includegraphics[width = 1.0\linewidth]{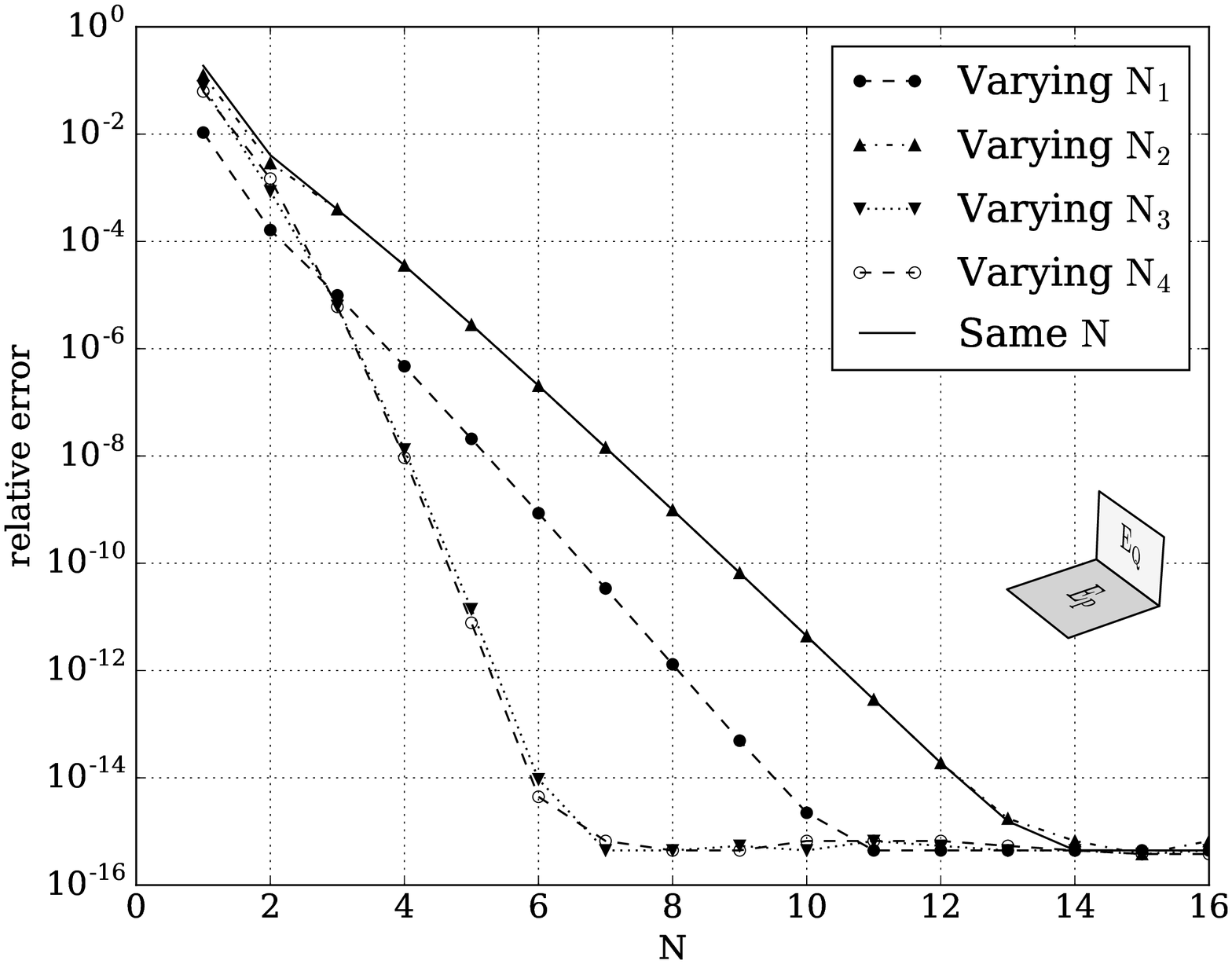}
	\caption{Relative error in computing the strongly singular integrals \eqref{I_mn_SS}  over edge-adjacent squares as a function of the order of the 1-D Gaussian quadrature.}
	\label{fig:error_SS_one_EAo}
\end{figure}

\begin{figure}[t!]
	\includegraphics[width = 1.0\linewidth]{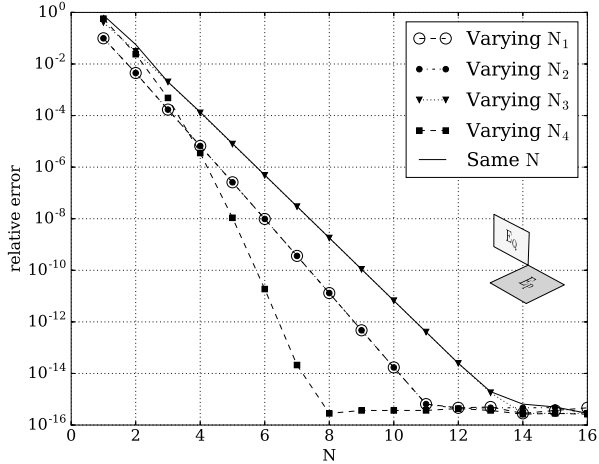}
	\caption{Relative error in computing the  strongly singular integrals \eqref{I_mn_SS}  over vertex-adjacent squares as a function of the order of the 1-D Gaussian quadrature.}
	\label{fig:error_SS_one_VAo}
\end{figure}

In the second set of experiments, the following weakly and strongly singular integrals are computed:
\begin{equation}
\label{I_mn_WS}
I^{\rm WS}_{m,n} = \int\limits_{E_P} \vec f_m(\vec r) \cdot\int\limits_{E_Q}  G(\vec r, \vec r') \cdot\vec f'_n(\vec r') dS' dS,
\end{equation}
\begin{equation}
\label{I_mn_SS}
I^{\rm SS}_{m,n} = \int\limits_{E_P} \vec f_m(\vec r) \cdot\int\limits_{E_Q} (\nabla G(\vec r, \vec r') \times \vec f'_n(\vec r')) dS' dS,
\end{equation}
where $E_P$ and $E_Q$ are observation and source quadrilateral elements, respectively. Here $\vec f_m(\vec r )$ and $\vec f_n (\vec r'), (m,n = 1,2,3,4)$ are vector basis functions of the 1\textsuperscript{st} order~\cite{Jin2014,Djordjevic2004}. Again, we consider the three singular integrals with coincident, edge adjacent and vertex adjacent patches. All patches are rectangular with edge-length equal to $d = 0.1\lambda$. The choice of the order of the quadrature rule for the associated 1-D integrals used in the previous example is by no means optimal. Hence, this time we vary the order of the integration rule for each one of the 1-D integrals while keeping the other three fixed and equal to $N=20$. The reference values are obtained by using a high
number of integration points for all four one-dimensional integrations, i.e., $N_1 = N_2 = N_3 = N_4 = 20$.
Finally, we evaluate the maximum relative error, defined as 
\[
\varepsilon_{max} = \max_{m,n = 1,2,3,4} \varepsilon_{m,n},
\]
where
\[
\varepsilon_{m,n} = \left\|{\frac{\|I_{m,n} - I_{m,n}^{\mathrm{ref}}\|_2}{\|I_{m,n}^{\mathrm{ref}}\|_2} + \epsilon }\right\|_2.
\]
As clearly shown in Figs.~\ref{fig:error_WS_one_ST}--\ref{fig:error_SS_one_VAo}, a predefined accuracy can be achieved with less integration points, hence, the efficiency of DIRECTFN-quad can be significantly improved by a judicious choice of integration orders for the different 1-D integrals. The optimal choice of the various integration orders is left for future work. 


\subsection{Singular integrals over quadratic curvilinear quadrilaterals}

\begin{figure}[t!]
	\centering\includegraphics[width = 1\linewidth]{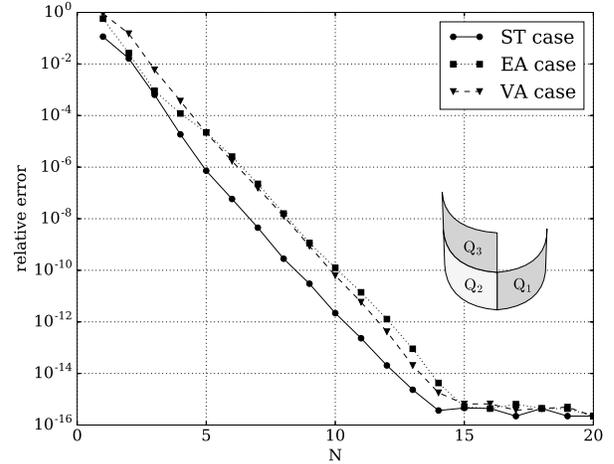}
	\caption{Relative error in computing the weakly and strongly singular integrals over quadratic curvilinear elements as a function of the order of the 1-D Gaussian quadrature}
	\label{fig:curv}
\end{figure}

In the third and last set of experiments the case of quadratic curvilinear elements, i.e. 9-node generalized quadrilaterals, is presented. The only difference from the algorithm for planar elements is in the surface parametrization, as described in literature~\cite{Kolundzija_book}. The weakly singular integrals~\eqref{I_mn_WS} are computed for $E_P \equiv E_Q \equiv Q_1$, corresponding to ST case, and strongly singular integrals~\eqref{I_mn_SS} are computed  for $E_P \equiv Q_1$, $E_Q \equiv Q_2, Q_3$, corresponding to EA and VA cases, respectively. The geometrical details of the curvilinear elements can be found in~\cite{DIRECTFN}. As illustrated in Fig.~\ref{fig:curv}, the fully numerical method presented herein can successfully handle the weakly and strongly singular integrals arising in Galerkin SIE formulations over curvilinear quadrilateral elements, without the need of modifying the main algorithms.

%% file: conclusion.tex
\section{CONCLUSION}

A collection of fully-numerical schemes is presented for the highly accurate and efficient evaluation of both weakly singular and strongly singular integrals arising from Galerkin surface integral equation methods for quadrilateral tessellations. Following the same rationale as in the case of triangular elements, the proposed method employs a series of variable transformations for the cancellation of the associated singularities. A key advantage of the novel algorithms is the further regularization of the integrands by means of appropriate re-ordering of the integrations. The resulting kernels of the four-dimensional integrals are sufficiently smooth functions with respect to all variables involved, thus allowing the use of simple Gauss quadrature rules. Finally, we note that the final algorithms are available as free, open-source software, readily applicable to a wide range of challenging cases, including weakly and strongly singular kernels, basis and testing functions of arbitrary order, planar and curvilinear patches, and problem-specific Green functions.